\documentclass[submission,copyright]{eptcs}
\providecommand{\event}{QFM 2012} 
\usepackage[english]{babel}
\usepackage[utf8]{inputenc}
\usepackage[T1]{fontenc}
\usepackage{tikz}
\usepackage{amssymb}
\usepackage[fleqn]{amsmath}

\DeclareMathAlphabet{\mathcal}{OMS}{cmsy}{m}{n}
\DeclareSymbolFont{letters}{OML}{txmi}{m}{it}

\usetikzlibrary{arrows}

\newcommand{\RN}{Radon-Nikod\'ym}

\newcommand{\keyword}[1]{\textsf{\textbf{#1}}}
\newcommand{\syntax}[1]{\textsf{#1}}
\newcommand{\var}[1]{\textsl{#1}}

\newcommand{\Sprobe}[1]{\var{Probe}~#1}
\newcommand{\Sstart}{\var{Start}}
\newcommand{\Sok}{\var{Ok}}
\newcommand{\Serror}{\var{Error}}

\newcommand{\Nat}{\ensuremath{\mathbb{N}}}
\newcommand{\Real}{\ensuremath{\mathbb{R}}}
\newcommand{\Ereal}{\ensuremath{\overline{\mathbb{R}}}}
\newcommand{\Bool}{\ensuremath{\mathbb{B}}}

\newcommand{\Kif}{\syntax{if}}
\newcommand{\Kelse}{\syntax{else}}
\newcommand{\Kthen}{\syntax{then}}
\newcommand{\Kcase}{\syntax{case}}
\newcommand{\Kof}{\syntax{of}}
\newcommand{\Kleast}{\syntax{LEAST}}

\newcommand{\Klemma}{\keyword{lemma}}
\newcommand{\Ktheorem}{\keyword{theorem}}

\newcommand{\Kdatatype}{\keyword{datatype}}

\newcommand{\Kand}{\keyword{and}}
\newcommand{\Kfixes}{\keyword{fixes}}
\newcommand{\Kassumes}{\keyword{assumes}}

\newcommand{\Tset}[1]{#1~\textsl{set}}

\newcommand{\Null}{\textbf{0}}

\newcommand{\Perr}{\ensuremath{P_{\var{err}}}}
\newcommand{\Cfin}{\ensuremath{C_{\var{fin}}}}

\newcommand{\jondos}{\var{jondos}}
\newcommand{\colls}{\var{colls}}
\newcommand{\init}{\var{init}}

\newcommand{\qAE}[1]{\textrm{A}\!\textrm{E}_{#1}\,}

\newcommand{\Kreachable}{\var{reachable}}
\newcommand{\Kuntil}{\var{until}}
\newcommand{\Kfair}{\var{fair}}
\newcommand{\Khitting}{\var{hitting-time}}
\newcommand{\Kfinite}{\var{finite}}
\newcommand{\Kmarkovchain}{\var{markov-chain}}
\newcommand{\Krewardedmarkov}{\var{markov-reward-chain}}

\title{Interactive verification of Markov chains: \\
Two distributed protocol case studies}
\author{
  Johannes Hölzl\thanks{Supported by the DFG Graduiertenkolleg 1480 (PUMA) and DFG project NI 491/10-2.}
  \quad\quad
  Tobias Nipkow  
  \institute{Technische Universit\"at München}
  \email{\quad \url{http://www.in.tum.de/~hoelzl} \quad\qquad \url{http://www.in.tum.de/~nipkow}}
}

\def\titlerunning{Interactive verification of Markov chains}
\def\authorrunning{J. Hölzl, T. Nipkow}

\begin{document}
\maketitle

\begin{abstract}
  Probabilistic model checkers like PRISM only check probabilistic systems of a
  fixed size. To guarantee the
  desired properties for an arbitrary size, mathematical analysis is necessary.
  We show for two case studies how this can be done in the interactive proof
  assistant Isabelle/HOL. The first case study is a detailed description
  of how we verified properties of the ZeroConf protocol, a decentral address
  allocation protocol. The second case study shows the more involved
  verification of anonymity properties of the Crowds protocol, an anonymizing
  protocol.
\end{abstract}

\section{Introduction}

The predominant approach to verification of probabilistic systems is model
checking~\cite{baier2008modelchecking}, and the most popular model checker is
PRISM~\cite{kwiatkowska2011prism4}. Model checking is automatic, but restricted to fixed
finite models. In this paper we put forward interactive theorem proving as a
realistic alternative approach that can deal with infinite-state systems on
an abstract mathematical level of Markov chains.  The specific contributions
of this paper are two case studies that illustrate our approach: the ZeroConf
protocol for decentralized address allocation and the anonymizing Crowds
protocol. The verifications are carried out in the proof assistant
Isabelle/HOL~\cite{nipkow2002isabelle}.

The characteristics of the theorem proving approach are:
\begin{itemize}
\item It can deal with infinite-state systems, although this paper considers only parameterized finite-state systems.

\item It is not restricted to some fixed set of concepts but user-extensible.

\item Logical soundness of the system depends only on the soundness of
a small fixed and trustworthy kernel of the theorem prover.

\item It requires familiarity with a theorem prover and a problem-dependent amount of work for each verification.
\end{itemize}
In a nutshell, it is mathematics, but checked by a computer.  These
characteristics indicate that the approach is more suitable for a research
environment than a product development environment.

\section{Formalization of probability in Isabelle/HOL}

To reason about Markov chains, especially about the probability that a path is
in a certain set, requires measure and probability theory. This section
gives a short introduction into the formalization of the theories required by
this paper. For a more detailed overview of the measure space formalization see
Hölzl and Heller~\cite{hoelzl2011measure}, and for the formalization of Markov
chains see Hölzl and Nipkow~\cite{hoelzl2012verifyingpctl}.

\subsection{Isabelle/HOL notation}

Isabelle/HOL largely follows ordinary mathematical notation. With a few
exceptions, we follow Isabelle/HOL notation in this paper, to give the reader
a better impression of the look-and-feel of the work. HOL is based on
$\lambda$-calculus. Hence functions are usually curried ($\tau_1 \to \tau_2
\to \tau_3$ rather than $\tau_1 \times \tau_2 \to \tau_3$) and function
application is written $f\;a$ rather than $f(a)$. The letters $\alpha$ and
$\beta$ stand for type variables. Type $\Bool$ is the type of boolean values.
Type $\tau\;set$ is the type of sets with elements of type $\tau$. Notation
$t :: \tau$ means that $t$ is a term of type $\tau$.  We regard functions of
type $\Nat\to\tau$ as infinite sequences of elements of type $\tau$.
Prepending an element $a::\tau$ to a sequence $\omega :: \Nat\to\tau$ is
written $a \cdot \omega$ and means $\lambda
i.~\Kif~i=0~\Kthen~a~\Kelse~\omega\;(i-1)$. The term $\Kleast\;n.~P\;n$ is
the least natural number $n$ such that $P\;n$ holds. If there is no such $n$,
then the term has some arbitrary (defined!) value, but we do not know which.

\subsection{Probability space} 

In this paper we are only interested in probabilities, hence we write
measures as $\Pr_s :: \Tset{\alpha} \to \Real$, where $s$ indicates the
particular probability measure under consideration. Similarly for the measurable
sets we write
$\mathcal{A}_s :: \Tset{\Tset{\alpha}}$ and for the entire space we write
$\Omega_s :: \Tset{\alpha}$. Here $\alpha$ is an arbitrary type where we cut
out a space $\Omega_s$. This is necessary as in many cases we are only
interested in a subset of the entire type, e.g.\ $\alpha$ is the type of
natural numbers $\Nat$ and we want to have a distribution on the finite subset
$\Omega_s = \{ 0, \ldots, N \}$. We usually drop $\Omega_s$ and write
$\{\omega \mid P~\omega\}$ instead of $\{\omega \in \Omega_s \mid P~\omega\}$
and $\Pr_s(\omega.\; P~\omega)$ instead of $Pr_s~\{ \omega \in \Omega_s.\; P~\omega \}$.

The measurable sets $\mathcal{A}_s$ form a $\sigma$-algebra, hence they are
closed under conjunction, disjunction, negation and countably bounded universal and existential quantification.
We have the defining
properties on the probability measure $\Pr_s$, as $\Pr_s~\emptyset = 0$,
$\Pr_s~\Omega_s = 1$, it is non-negative: $0 \le \Pr_s~A$ and
countably additive: For a measurable and disjoint family $P :: \Nat \to \alpha
\to \Bool$
\[ \textstyle \Pr_s(\omega.\; \exists i.\; P~i~\omega ) = (\sum_i \Pr_s(\omega.\; P~i~\omega)) \ . \] 

For a finite probability space measurable sets need only be closed under
finite bounded quantifiers, and the probability needs only be finitely
additive, instead of countably additive. Unfortunately, the path space on
Markov chains is neither finite nor discrete, so we need $\sigma$-algebras
and countably additive probability measures.

We also need conditional probability and define it as usual:
\[ \textstyle
\Pr_s(\omega.\; P~\omega \mid Q~\omega) =
\Pr_s(\omega.\; P~\omega \land Q~\omega) / \Pr_s(\omega.\; Q~\omega ) \ .
\]
%

The A\!E-quantifier $\qAE{s}\omega.\; P~\omega$ on a path measure $\Pr_s$ states
that the property $P$ holds with probability~$1$. Isabelle/HOL also has a
formalization of the Lebesgue integral on probability spaces, as notation we use
$\int_\omega f~\omega\;d\!\Pr_s$.


\subsection{Markov chains} 

We introduce Markov chains as probabilistic automata, i.e.\ as discrete-time
time-homogeneous finite-space Markov processes. A Markov chain is defined by its
state space $S :: \Tset{\alpha}$ and an associated transition matrix $\tau :: \alpha \to \alpha \to \Real$. We assume no
initial distribution or starting state, however when measuring paths we always
provide a starting state. A path on a Markov chain is a function $\Nat
\to S$, i.e.\ an infinite sequence of states visited in the Markov chain.

\[
  \Kmarkovchain~S~\tau =
    \Kfinite~S \land S \neq \emptyset \land
    \Big(\forall s, s' \in S. ~ 0 \leq \tau~s~s'\Big) \land
    \Big(\forall s \in S. ~\big(\sum_{s' \in S} \tau~s~s'\big) = 1\Big)
\]

For the rest of this section we assume a Markov chain with state space $S$ and
transition matrix $\tau$. We write $E(s)$ for the set of all successor states,
i.e.\ all $s' \in S$ with $\tau~s~s' \neq 0$. Note that a path $\omega$ does not
require that $\omega~(i+1)$ is a successor of $\omega~i$.

We have defined a probability space $(\Nat \to S, \mathcal{A}, \Pr_s)$ on
the space of all paths $\Nat \to S$ for a starting state $s \in S$. The
measurable sets are the $\sigma$-algebra generated by all sets $\{ \omega \in
\Nat\to S \mid \omega~i = t \}$ where $i \in \Nat$ and $t \in S$.
The measure $\Pr_s$ (depending on the starting state $s \in S$ and $\tau$)
is defined via an infinite product and is shown to satisfy the following
key property (where $s{\cdot}\omega$ prepends $s$ to $\omega$):
\[
\textstyle
\forall \omega \in \Nat \to S, s \in S, n.\,~
  \Pr_s~\{\omega' \mid \forall i<n.\; \omega'~i = \omega~i \} =
  \prod_{i<n} \tau~((s{\cdot}\omega)~i)~(\omega~i)
\]
Note that $\Pr_s$ explicitly carries the starting state
and yields the transition probability for the steps $s \to_\tau \omega~0 \to_\tau \omega~1
\to_\tau \dots \to_\tau \omega~(n-1)$.

We also use Markov reward chains, where we assign a cost or reward to each transitions:
\[
  \Krewardedmarkov~S~\tau~\rho = \Kmarkovchain~S~\tau \land (\forall s, s' \in S.\; 0 \le \rho~s~s')
\]

This approach allows a very easy definition of a Markov chain given as a
transition system. Other formalizations of Markov
chains~\cite{hurd02thesis,liu2011markovchains} use the probability space
$\Nat \to\ \Bool$. This
requires to provide a measurable function $X~t~\omega$, mapping a sequence of
boolean choices $\omega :: \Nat \to \Bool$ into a state at time $t$. In our
approach the set of states $S$ and the transition matrix $\tau$ are enough.

Some models require an arbitrary set $I$ of independent variables $X_i$
with distribution $P_i$. For this case we provide the product $\prod_I P_i$. We
use this product space to construct the path space for our Markov chains.
Furthermore the probability space $\Nat \to \Bool$ is just a special instance of
the generalized product space.

\subsubsection{Iterative equations}

The Markov chain induces \emph{iterative equations} on the probability $\Pr_s$,
the Lebesgue integral and the A\!E-quantifier, relating properties about $s$ to
properties of $E(s)$. These
equations are often useful in inductive proofs and already give a hint how to
prove concrete properties of probabilities and integrals. If $A$, $P$, and
$f$ are measurable and $s \in S$, then the following equations hold:
\[
\begin{array}{rcl}
 \Pr_s~A & = & \displaystyle
    \sum_{s' \in E(s)}~\tau~s~s' * \Pr{}_{s'}
    (\omega.\; s'{\cdot}\omega \in A) \\[1em]
\displaystyle    \int_{\omega} f~\omega\; d\!\Pr{}_s & = & \displaystyle
    \sum_{s' \in E(s)}~\tau~s~s' *
    \int_{\omega} f~(s'{\cdot}\omega)\; d\!\Pr{}_{s'} \\[1.5em]
\displaystyle    \qAE{s}\omega.~P~\omega & = & \displaystyle
    \forall s' \in E(s).~\qAE{s'}\omega.~P~(s'{\cdot}\omega)
\end{array}
\]

\subsubsection{Reachability}

Let $\Phi$ be a subset of $S$. A state $s'$ is \emph{reachable} via $\Phi$
starting in $s$ iff there is a non-zero probability to reach $s'$ by only
going through the specific set of states $\Phi$. The starting state $s$ and
the final state $s'$ need not be in $\Phi$.
\[ \begin{array}{l@{}l}
  \Kreachable~\Phi~s := \{s' \in S \mid \exists \omega \in \Omega, n.\; &
   (\forall i \leq n.~\omega~i \in E((s{\cdot}\omega)~i)) \ \land \\
  & (\forall i < n.~\omega~i \in \Phi) \land \omega~n = s' \}
\end{array} \]
Reachability is a purely qualitative property, as it is defined on the graph of
non-zero transitions.

The until-operator introduces a similar concept on paths. Its definition does
not assume that a state is a successor state of the previous one, as this is
already ensured by~$\Pr_s$.
\[ \Kuntil~\Phi~\Psi = \{\omega \mid
   \exists n.~(\forall i < n.~\omega~i \in \Phi) \land \omega~n \in \Psi\} \]
Can we compute $\Pr_s (\Kuntil~\Phi~\Psi)$ using only
$\Kreachable$? It is easy to show that $\Pr_s (\Kuntil~\Phi~\Psi) = 0$ iff
$(\Kreachable~\Phi~s) \cap \Psi = \emptyset$. But is there also a way to
characterize $\Pr_s (\Kuntil~\Phi~\Psi) = 1$ in terms of $\Kreachable$?

\subsubsection{Fairness}
   
To show that $\Kreachable$ can be used to guarantee that states are reached with
probability $1$, we need state fairness. A path $\omega$ is
\emph{state fair} w.r.t.\ $s$ and $t$ if $s$ appears only finitely often provided
that $t$ also appears only finitely often as the successor of $s$ in $\omega$. The
definition and proofs about state fairness are based on the thesis by
Baier~\cite{baier1998habil}.
\[
  \Kfair~s~t = \big\{ \omega \mid
     \Kfinite~\{ n \mid \omega~n = s \land \omega~(n+1) = t \}
     \implies \Kfinite~\{ n \mid \omega~n = s \} \big\}
\]
We show that almost every path is state fair for each state and its successors.
\[ \forall s, s' \in S, t' \in E(s').~
  \qAE{s}\omega.~ s{\cdot}\omega \in \Kfair~s'~t'
\]

Using this we prove that starting in a state $s$ almost every path fulfills
$\Kuntil~\Phi~\Psi$ if (1) all states reachable via $\Phi$ are in $\Phi$ or
$\Psi$ and (2) each state reachable from $s$ has the possibility to reach
$\Psi$. This theorem allows us to prove that $\Kuntil~\Phi~\Psi$ holds almost
everywhere by a reachability analysis on the graph:
\[ \begin{array}{l}
s \in \Phi \ \land \ \Phi \subseteq S \ \land \ 
\Kreachable~(\Phi \setminus \Psi)~s \subseteq \Phi \cup \Psi \ \land \\
\forall t \in (\Kreachable~(\Phi \setminus \Psi)~s \cup \{s\}) \setminus \Psi.~
\Kreachable~(\Phi \setminus \Psi)~t \cap \Psi \neq \emptyset \\
\implies \qAE{s}\omega.~s{\cdot}\omega \in
\Kuntil~\Phi~\Psi
\end{array} \]

\subsubsection{Hitting time}

The \emph{hitting time} on a path $\omega$ is the first index at which a state from a set $\Phi$ occurs:
\[ \Khitting~\Phi~\omega = \Kleast~i.\; \omega~i \in \Phi \]
Note that if there is no $i$ such that $\omega~i \in \Phi$, then
$\Khitting~\Phi~\omega$ is some arbitrary, underspecified natural number.
For the computation of rewards it is important to know if the expected hitting
time is finite. We
show that the expected hitting time of $\Phi$ for paths starting in $s$ is
finite if almost every path starting in $s$ reaches $\Phi$. If $s$ is in $S$ and
$\qAE{s}\omega.~s{\cdot}\omega \in \Kuntil~S~\Phi$ then
\[ \int_\omega \Khitting~\Phi~(s{\cdot}\omega) d\!\Pr{}_s \neq \infty \]

For Markov reward chains we are interested in the transition costs until a set
of states occurs:
\[ \textstyle
\var{cost-until}~\Phi~\omega = \Kif~\exists i.\; \omega~i \in \Phi~\Kthen~\sum_{i < \Khitting~\Phi~\omega} \rho~(\omega~i)~(\omega~(i + 1)) ~\Kelse~\infty \]
\section{Case study: The ZeroConf protocol}

Ad-hoc networks usually do not have a central address authority assigning
addresses to new nodes in the network. An example are consumer networks where
users want to connect their laptops to exchange data or attach a network capable
printer. When connecting with WiFi these devices use IPv4 and hence need IPv4
addresses to communicate with each other.

The \emph{ZeroConf} protocol
\cite{cheshire2005zeroconf} is a distributed network protocol which
allows new hosts in the network to allocate an unused link-local IPv4 address. A
link-local address is only valid in the local network, e.g.\ a WiFi
network. We assume point-to-point communication in our local
network, and hence communicate directly with each host identified by a valid
address. The problem with IPv4 addresses is that they are limited, i.e.\ they
are represented by 32-bit numbers, and for the local network the addresses from
$169.254.1.0$ to $169.254.254.255$ are available, hence we can chose from
$65024$ distinct addresses. ZeroConf works by randomly selecting an address from
this pool and then probing if the address is already in use.

Bohnenkamp~\emph{et~al.}~\cite{bohnenkamp2003zeroconfcost} give a formal analysis of
the probability that an address collision happens, i.e.\ two hosts end up with
the same address. They also analyse the expected run time until a (not necessaryly
valid) address is chosen. As our first case study we formalize their analysis in
Isabelle/HOL.

Andova~\emph{et~al.}~\cite{andova2003rewardsmodel} present a model-checking
approach for discrete-time Markov reward chains and apply it to the ZeroConf
protocol as a case study. They support multiple reward structures and can
compute the probability based on multiple constraints on these reward
structures. Kwiatkowska~\emph{et~al.}~\cite{KNPS06} have modelled this protocol
as a probabilistic timed automata in PRISM. Both models include more features of
the actual protocol than the model by
Bohnenkamp~\emph{et~al.}~\cite{bohnenkamp2003zeroconfcost} that we follow.

\subsection{Description of address allocation} 

We give a short description of the model used in
Bohnenkamp~\emph{et~al.} \cite{bohnenkamp2003zeroconfcost}.
The address allocation in ZeroConf uses ARP (address resolution protocol) to
detect if an address is in use or not. An ARP request is sent to detect if a
specific IPv4 address is already in use. When a host has the requested IPv4
address it answers with an ARP response.
ZeroConf allocates a new address as follows:

\begin{enumerate}
  \item Select uniformly a random address in the range $169.254.1.0$ to
    $169.254.254.255$.
  
  \item Send an ARP request to detect if the address is already in use.
  
  \item When a host responds to the ARP request, the address is already taken
    and we need to start again (go back to 1).
    
  \item When no response arrives before a time limit $r$, we again send
    an ARP request. This is repeated $N$ times.
    
  \item When no response arrived for $N$ requests we assume our address is not
    in use and are finished.
\end{enumerate}

This probabilistic process depends on two parameters: (1) The probability $q$
that the random chosen address is already taken; this probability depends on the number
of hosts in the network and the number of available addresses. (2) The
probability $p$ that either the ARP request or response is lost.

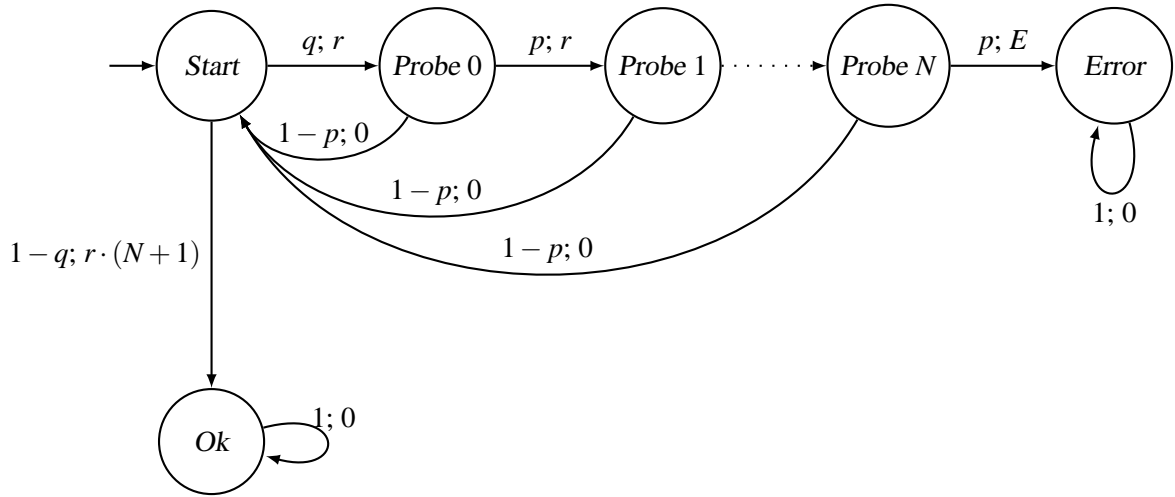
\begin{figure} 
\begin{tikzpicture}[thick]

  \node (bot)   at (-1.5, 0) {} ;

  \node[draw,circle] (start) at ( 0,  0) {$~~\Sstart~~$} ;

  \node[draw,circle] (probe0)  at ( 3,  0) {$\Sprobe{0}$} ;

  \node[draw,circle] (probe1)  at ( 6,  0) {$\Sprobe{1}$} ;

  \node[draw,circle] (probeN)  at ( 9,  0) {$\Sprobe{N}$} ;
      
  \node[draw,circle] (error)  at (12,  0) {$~~\Serror~~$} ;

  \node[draw,circle] (ok)  at ( 0, -5) {$~~~\Sok~~~$} ;

  \path[->, >=latex] (bot)   edge (start)
            (start)  edge node [above] {$q$; $r$} (probe0)
                     edge node [left]  {$1 - q$; $r \cdot (N + 1)$} (ok) 
            (probe0) edge node [above] {$p$; $r$} (probe1)
                     edge [bend left=60] node [above] {$1 - p$; $0$} (start)
            (probe1) edge [loosely dotted] (probeN)
                     edge [bend left=60] node [above] {$1 - p$; $0$} (start)
            (probeN) edge [bend left=60] node [above] {$1 - p$; $0$} (start)
                     edge node [above] {$p$; $E$} (error)
            (ok)     edge [loop right] node [above] {~~$1$; $0$} (ok)
            (error)  edge [loop below] node [below] {$1$; $0$} (error);
  
\end{tikzpicture}

\caption{Markov chain of the ZeroConf~protocol. The labels are annotated with
  $P; T$: the probability $P$ to take this edge and the elapsed time $T$.}
\label{fig:zeroconf}
\end{figure}

The Markov chain shown in Fig.~\ref{fig:zeroconf} describes the address
allocation from a global viewpoint. At \Sstart{} a new host is added to the
network, it chooses an address and sends the first ARP request.
There are two alternatives.
\begin{itemize}
\item With probability $1 - q$ the host chooses an unused address, the
  allocation is finished, the Markov chain directly goes to \Sok{}. Of
  course, the host does not know this, and still sends out $N + 1$ ARP
  probes. Hence we associate the time cost $r \cdot (N + 1)$ with this
  transition.
\item With probability $q$ the host chooses a used address and
  goes to the probing phase: In the \Sprobe{$n$}{} state it sends an ARP
  request and waits until $r$ time units have passed, or until it receives an ARP response from
  the address owner. With probability $1 - p$ the host receives an ARP
  response and needs to choose a new address---we go back to \Sstart{}.  With
  probability $p$ this exchange fails and we go to the next probe phase.
  After $N + 1$ probes, the host assumes the chosen address is free. As two
  hosts in the network end up with the same address we reached the \Serror{}
  state. The time cost $E$ models the cost to repair the double allocation.
  This might involve restarting a laptop.
\end{itemize}

\subsection{Formal model of ZeroConf address allocation} 

The Isabelle/HOL model of the ZeroConf~protocol describes the Markov chain in
Fig.~\ref{fig:zeroconf}. We set up a context containing the probe numbers (starting
with $0$), the probabilities $p$ and $q$, and the costs $r$ and $E$:
\[ 
  \begin{array}{@{}l}
        \Kfixes~ N :: \Nat ~\Kand~ p~q~r~E :: \Real \\
     \Kassumes~ 0 < p ~\Kand~ p < 1 ~\Kand~ 0 < q ~\Kand~ q < 1 \\
     \Kassumes~ 0 \le E ~\Kand~ 0 \le r
  \end{array}
\] 
In the following sections we assume that these fixed variables $N$, $p$, $q$,
$r$, and $E$ fulfill the above assumptions of the ZeroConf protocol.

To represent the states in the Markov chain we introduce a new datatype:
\[ 
  \Kdatatype~\var{zc-state} = \Sstart \mid \Sprobe{\Nat} 
    \mid \Sok \mid \Serror 
\] 
We have the type \var{zc-state} with the distinct objects \Sstart, \Sok, \Serror,
and $\Sprobe{n}$ for all $n :: \Nat$. The valid states $\var{S} :: \Tset{\var{zc-state}}$ are a
restriction of this to only valid probe numbers. This also gives us a finite
number of states.
\[ 
  \var{S} = \left\{ \Sstart, \Sok, \Serror \right\}
    \cup \left\{ \Sprobe{n} \mid n \le N \right\}
\] 
The final modeling step is to define the transition matrix $\tau ::
\var{zc-state} \to \var{zc-state} \to \Real$ and the cost function $\rho ::
\var{zc-state} \to \var{zc-state} \to \Real$. Both are defined by a case
distinction on the current state and return the zero function \Null{} which is
updated at the states with non-zero transition probability or cost.
\[ 
  \begin{array}{@{}r@{~}l@{~}l}
  \tau~s = \Kcase~s~
    \Kof&\Sstart & \Rightarrow \Null(\Sprobe{0} := q, \Sok := 1 - q) \\ 
  \mid &\Sprobe{n} & \Rightarrow
      \Kif~n < N~\Kthen~\Null(\Sprobe{(n+1)} := p, \Sstart := 1 - p) \\
  & & \hphantom{\Rightarrow \Kif~n < N~} \Kelse~\;\, \Null(\Serror := p, \Sstart := 1 - p) \\
  \mid &\Sok & \Rightarrow \Null(\Sok := 1) \\ 
  \mid &\Serror & \Rightarrow \Null(\Serror := 1) \\
  \\
  \rho~s = \Kcase~s~
    \Kof & \Sstart & \Rightarrow \Null(\Sprobe{0} := r, \Sok := r * (N + 1)) \\ 
  \mid & \Sprobe{n} & \Rightarrow
      \Kif~n < N~\Kthen~\Null(\Sprobe{(n+1)} := r) ~\Kelse~ \Null(\Serror := E) \\
  \mid & \Sok & \Rightarrow \Null \\ 
  \mid & \Serror & \Rightarrow \Null \\ 
  \end{array}
\] 

We need to prove that we actually defined a Markov chain: as a consequence,
Isabelle/HOL is able to provide the probabilities $\Pr_{s} A$ for each
state $s$ and
path set $A$. For this we show that $\tau$ is a valid transition matrix for
a Markov chain on $S$, and $\rho$ is a valid cost function:
\[ \Ktheorem~\text{$\tau$-DTMC:}\qquad \Krewardedmarkov~S~\tau~\rho \]
To prove this we need to show that $\tau$ and $\rho$ are non-negative for all
states in $S$. And finally we need to show that $\tau~s$ is a distribution for
all $s$ in $S$, which is easy to show by using the helper lemma $S$-split:
\[ \Klemma~\text{$S$-split:}\qquad \sum_{s \in S} f~s =
  f~\Sstart + f~\Sok + f~\Serror + \sum_{n \le N} f~(\Sprobe{n})\]

  
\subsection{Probability of an erroneous allocation} 

The correctness property we want to verify is that no collision happens, i.e.\
we want to compute the probability that a protocol run ends in the \Serror{}
state. The goal of this section is not only to show \emph{what} we proved, but
to show \emph{how} we proved it. Most of the proofs are automatic by
rewriting and we do not show the details. But we want to show the necessary
lemmas and theorems needed to convince Isabelle/HOL. 

We define $\Perr{} :: \var{zc-state} \to \Real$ to reason about the probability
that a trace $\omega$ ends in the \Serror{} state when we started in a state
$s$:
\[ \Perr~s =
   \Pr{}_s\big(\omega.\;s {\cdot} \omega \in \var{until}~S~\{\Serror\}\big) \]
Our final theorem will be to characterize $\Perr~\Sstart$ only in terms of
the system parameters $p$, $q$ and $N$.

The first obvious result is that when we are already in \Serror{}, we will stay
in \Serror{}, and when we are in \Sok{} we will never reach \Serror{}:
\[
  \begin{array}{@{}ll}
    \Klemma~\text{\Perr-error:} & \Perr~\Serror = 1 \\
    \Klemma~\text{\Perr-ok:}\qquad & \Perr~\Sok = 0
  \end{array}
\]
\Perr-error is proved by rewriting: $\Serror \cdot \omega \in
\var{until}~S~\{\Serror\}$ is always true. The \Sok{} case is proved by
$\var{reachable}~(S \setminus \{\Serror\})~\Sok \subseteq \{\Sok\}$. Together
with lemma $S$-split and these two lemmas we provide an iterative lemma for \Perr{}:
\[ \begin{array}{@{}l}
  \Klemma~\text{\Perr-iter:}\\
  \qquad s \in S \implies \Perr~s = \tau~s~\Sstart * \Perr~\Sstart + \tau~s~\Serror + \sum_{n \le N} \tau~s~(\Sprobe{n}) * \Perr~(\Sprobe{n})
  \end{array}
\]
However this is a bad rewrite theorem, using it would result in
non-termination of the rewrite engine. To avoid this we derive rules for
specific states:   
\[
\begin{array}{@{}ll}
  \Klemma~\text{\Perr-last-probe:} & \Perr~(\Sprobe{N}) = p + (1 - p) * \Perr~\Sstart \\
  \Klemma~\text{\Perr-start-iter:}\qquad & \Perr~\Sstart = q * \Perr~(\Sprobe{0})\\
\end{array}
\]

Our next step is to compute the probability to reach \Serror{} when we are
in \Sprobe{$n$}. This is the only proof which is not done by a simple rewrite step,
but it requires induction and two separate rewrite steps. The induction is done
over the number $n$ of steps until we are in \Serror{}. To
give the reader a better feeling for what these proofs look like,
here is the skeleton of the Isabelle proof:
\[ \begin{array}{@{}l}
 \Klemma~\text{\Perr-probe-iter:}\qquad~n \le N \implies \Perr~(\Sprobe{(N - n)}) =
   p^{n + 1} + (1 - p^{n + 1}) * \Perr~\Sstart \\
  \keyword{proof}~(\var{induct}~n) \\
  \quad \keyword{case}~(n + 1) \\
  \quad \keyword{have}~\Perr~(\Sprobe{(N - (n + 1))}) = 
    p*(p^{n + 1} + (1 - p^{n + 1})*\Perr~\Sstart) + (1 - p)*\Perr~\Sstart\\
  \quad \quad \var{<proof>} \\
  \quad \keyword{also}~\keyword{have}~\dots = p^{(n + 1) + 1} + (1 - p^{(n + 1) + 1}) * \Perr~\Sstart \\
  \quad \quad \var{<proof>} \\
  \quad \keyword{finally}~\keyword{show}~\Perr~(\Sprobe{(N - (n + 1))}) = p^{(n + 1) + 1} + (1 - p^{(n + 1) + 1}) * \Perr~\Sstart \ .\\
  \keyword{qed}~\var{simp} \quad \text{-- The $0$-case is a simple rewriting step with \Perr-last-probe.}
\end{array} \] 
Together with \Perr-start-iter we prove our final theorem:
\[ \Ktheorem~\text{\Perr-start:}\qquad~\Perr~\Sstart = 
  (q * p^{N + 1}) / (1 - q * (1 - p^{N + 1})) \]

With typical parameters for the ZeroConf protocol (16 hosts
($q = 16 / 65024$), 3 probe runs ($N = 2$) and a probability of $p = 0.01$ to
lose ARP packets) we compute (by rewriting) in Isabelle/HOL that the probability to reach
\Serror{} is below $1 / 10^{13}$:
\[ \keyword{theorem}~\Perr~\Sstart \le 1 / 10^{13} \]
  

\subsection{Expected running time of an allocation run} 

Users are not only interested in a very low error probability
but also in fast allocation time for network address.
Obviously there are runs which may take very long, but the probability for these
runs are near zero. So we want to verify that the average running time of an
allocation run is in the time range of milliseconds. 

The running time of an allocation run $\Cfin :: S \to \Ereal$ is modelled as the
integral over the sum of all costs $\rho$ for each step in each run. The sum of
all steps until either \Sok{} or \Serror{} is reached is simply
$\var{cost-until}$:
\[
\Cfin~s = \int_\omega \var{cost-until}~\{\Serror, \Sok\}~(s \cdot \omega)~\text{d}\!\Pr{}_s
\]
In order to evaluate the integral we first show that it is finite.
This is the case if $\var{cost-until}~\{\Serror, \Sok\}$ is finite
almost everywhere.
So we first show that almost every path reaches $\{\Serror, \Sok\}$:
\[
  \Klemma~\text{AE-term:}\qquad s \in S \implies
  \text{AE}_s\,\omega. \; s\cdot\omega \in \var{until}~S~\{\Serror, \Sok\}
\]
Using this we show an elementary form of \Cfin{} in a similar way to \Perr:   
\[
  \begin{array}{@{}l}
  \Klemma~\text{\Cfin-start:}\qquad \Cfin~\Sstart = \\
  \quad 
  (q * (r + p^{N + 1} * E + r * p * (1 - p^N) / (1 - p)) + (1 - q) * (r * {N + 1})) /
    (1 - q + q * p^{N + 1})
  \end{array}
\]

With typical values (16 hosts, 3 probe runs, a probability of $p = 0.01$ to
lose ARP packets, $2~ms$ for an ARP round-trip ($r = 0.002$) and an error
penalty of one hour ($E = 3600$)) we compute in Isabelle/HOL that the average time
to terminate is less or equal $0.007~s$:
\[ \keyword{theorem}\qquad\Cfin~\Sstart \le 0.007 \]


\section{Case study: The Crowds protocol}

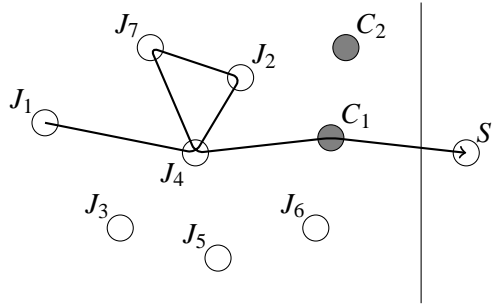
\begin{figure} 
\begin{center}
\begin{tikzpicture}

  \draw (0,    0.4) circle (1ex) node [above left] {$J_1$} ;
  \draw (1. , -1. ) circle (1ex) node [above left] {$J_3$} ;
  \draw (2.6, +1. ) circle (1ex) node [above right] {$J_2$} ;
  \draw (2  ,  0  ) circle (1ex) node [below left] {$J_4$} ;
  \draw (1.4,  1.4) circle (1ex) node [above left] {$J_7$} ;
  \draw (3.6, -1. ) circle (1ex) node [above left] {$J_6$} ;
  \draw (2.3, -1.4) circle (1ex) node [above left] {$J_5$} ;
  
  \draw[fill=gray] (3.8,  0.2) circle (1ex) node [above right] {$C_1$} ;
  \draw[fill=gray] (4.0,  1.4) circle (1ex) node [above right] {$C_2$} ;
  
  \draw (5.6,  0  ) circle (1ex) node [above right] {$S$} ;

  \draw (5.0, 2) -- (5.0, -2) ;
  
  \draw[->, thick, rounded corners=3pt] (0, 0.4) -- (2, 0) -- (2.6, +1.) -- (1.4, 1.4) -- (2, 0) -- (3.8, 0.2) -- (5.6, 0) ;
  
\end{tikzpicture}
\end{center}
\caption{The established route $J_1 - J_4 - J_2 - J_7 - J_4 - C_1 - S$}
\label{fig:crowds}
\end{figure} 

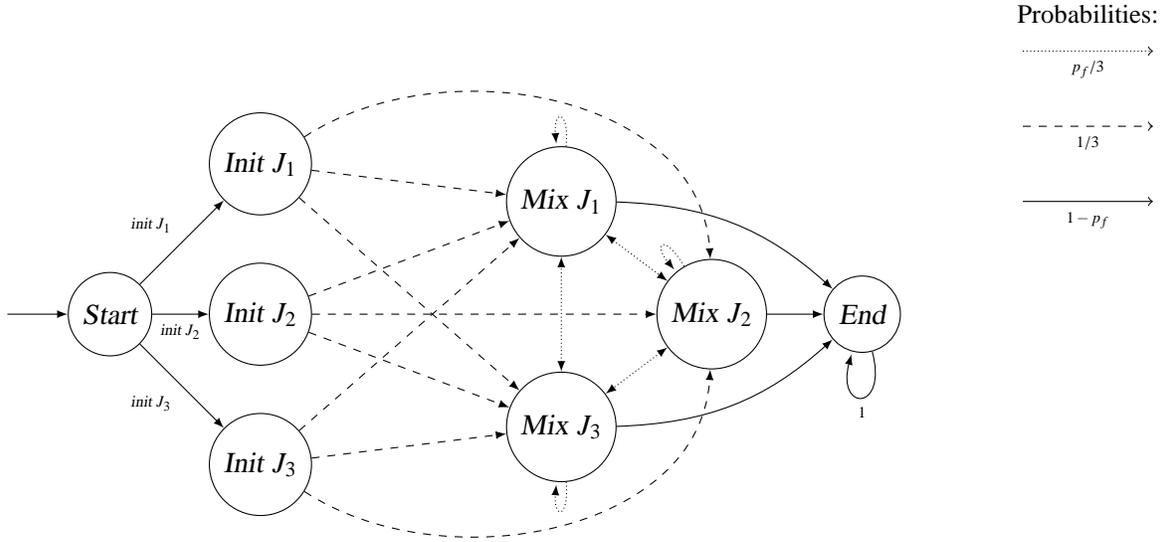
\begin{figure} 
\begin{center}
\begin{tikzpicture}

\node (bot)   at (-1.5, 0) {} ;

\node[draw, circle] (S) at (0,  0) {$\var{Start}$} ;

\node[draw, circle] (J1) at (2,  2) {$\var{Init}~J_1$} ;
\node[draw, circle] (J2) at (2,  0) {$\var{Init}~J_2$} ;
\node[draw, circle] (J3) at (2, -2) {$\var{Init}~J_3$} ;
                                   
\node[draw, circle] (M1) at (6,  1.5) {$\var{Mix}~J_1$} ;
\node[draw, circle] (M2) at (8,  0) {$\var{Mix}~J_2$} ;
\node[draw, circle] (M3) at (6, -1.5) {$\var{Mix}~J_3$} ;
                                   
\node[draw, circle] (E) at (10,  0) {$\var{End}$} ;

\path[->, >=latex]
  (bot) edge (S)
  (S) edge node [above left] {\tiny $\var{init}~J_1$} (J1) 
      edge node [below] {\tiny $\var{init}~J_2$} (J2) 
      edge node [below left] {\tiny $\var{init}~J_3$} (J3) ;
      
\path[->, >=latex, dashed]
  (J1) edge (M1)
       edge [bend left=50, in = 110] (M2) 
       edge (M3)
  (J2) edge (M1)
       edge (M2) 
       edge (M3)
  (J3) edge (M1)
       edge [bend left=-50, in=-110] (M2) 
       edge (M3) ;
       
\path[->, >=latex]
  (M1) edge [bend left=20] (E)
  (M2) edge (E)
  (M3) edge [bend left=-20] (E) ;
  
\path[->, >=latex]
  (E)  edge [loop below] node [below] { \tiny $1$ } (E);

\path[->, >=latex, densely dotted]
  (M1) edge [in=95,out=85,loop] (M1)
  (M2) edge [in=130,out=120,loop] (M2)
  (M3) edge [in=-95,out=-85,loop] (M3) ;
  
\path[<->, >=latex, densely dotted]
  (M1) edge (M2)
  (M2) edge (M3)
  (M3) edge (M1) ;

\node (label) at (13, 4) { \small Probabilities: };

\node (s1) at (12, 3.5) {};
\node (t1) at (14, 3.5) {};
\path[->, densely dotted] (s1) edge node [below] {\tiny $p_f / 3$} (t1);

\node (s2) at (12, 2.5) {};
\node (t2) at (14, 2.5) {};
\path[->, dashed] (s2) edge node [below] {\tiny $1 / 3$} (t2);

\node (s3) at (12, 1.5) {};
\node (t3) at (14, 1.5) {};
\path[->] (s3) edge node [below] {\tiny $1 - p_f$} (t3);

\end{tikzpicture}
\end{center}
\caption{Example Markov chain of the small Crowds network $\{ J_1, J_2, J_3 \}$}
\label{fig:crowds-ex-mc}
\end{figure} 

The \emph{Crowds} protocol described by Reiter and Rubin~\cite{reiter1998crowds}
is an anonymizing protocol. The goal is to allow users to
connect to servers anonymously. Neither the final server should know which user connects to
it, nor attackers collaborating in the network. The Crowds protocol establishes
an anonymizing route through a so called mix network: Each user (Reiter and
Rubin name them \emph{jondo} pronounced ``John Doe'') is itself participating in
the mix network. When a jondo establishes a route, it first connects to another
random jondo which then decides based on a coin flip weighted with $p_f$ if it
should connect to the final server, or go through a further jondo, and so on.
Figure~\ref{fig:crowds} shows an established route through the jondos $J_1 - J_4
- J_2 - J_7 - J_4 - C_1 - S$. There is no global information about a route
available to the participating jondos. For each connection a jondo only knows
its immediate neighbours, but no other previous or following jondo, so it may
happen that a route is going through a loop, as seen in Fig.~\ref{fig:crowds}.

First, Reiter and Rubin~\cite{reiter1998crowds} show that the server has no
chance to guess the original sender. In a second step they assume that some
jondos collaborate to guess the jondo initiating the route. They analyse the
probability that a collaborating node is the successor of the initiating jondo.
This analysis is affected by the fact that the route may go through the
initiating jondo multiple times.
An analysis of the Crowds protocol in PRISM, for specific sizes, has been
conducted by Shmatikov~\cite{Shmatikov04}.

Similar to the ZeroConf case, we only analyse the Markov chain having a
global view on the protocol. We could model the individual behaviour of jondos in
Isabelle/HOL and show that this induces our Markov chain model, but this is not in
the scope of this paper.

\subsection{Formalization of route establishment in the Crowds protocol}

We concentrate on the probabilistic aspects of route establishment in the
Crowds protocol. We assume a set \jondos{} of an arbitrary type $\alpha$ (which
is just used to uniquely identify jondos), and a strict subset \colls, the
collaborating attackers. A jondo decides with probability $p_f$ if it
chooses another jondo as next step, or if it connects directly to the server.
The distribution of the initiating jondos is given by $\var{init}$. Naturally
the initiating jondo is not a collaborating jondo. In Isabelle this is expressed as the following context:
\[ 
\begin{array}{@{}l}
  \Kfixes~ \jondos~\var{colls} :: \Tset{\alpha} ~\Kand~ p_f :: \Real
        ~\Kand~ \init :: \alpha \to \Real\\
  \Kassumes~ 0 < p_f ~\Kand~ p_f < 1 \\
  \Kassumes~ \jondos \not= \emptyset ~\Kand~ \colls \not= \emptyset ~\Kand~ \var{finite}~\jondos
      ~\Kand~ \colls \subset \jondos \\
  \Kassumes~ \forall j \in \jondos.\;0 \le \init~ j ~\Kand~
      \forall j \in \colls.~\init~j = 0 ~\Kand~
      \sum_{j \in \jondos} \init~j = 1
  \end{array}
\] 

The Markov chain has four different phases: start, the initial node, and the
mixing phase, and finally the end phase where the server is contacted. See 
Fig.~\ref{fig:crowds-ex-mc} for a small example. Our
formalization of Markov chains requires a single start node, otherwise we could
choose \var{init} as initial distribution. The type of the state
$\alpha~\var{c-state}$ depends on the type of the jondos $\alpha$.
\[ 
\Kdatatype~\alpha~\var{c-state} = \var{Start} \mid
  \var{Init}~\alpha \mid \var{Mix}~\alpha \mid \var{End}
\] 
Similar to the ZeroConf protocol not all possible values of $\var{c-state}$ are
necessary. We restrict them further by only allowing non-collaborating jondos as
initial jondos, and only elements from \jondos{} participate in the mixing
phase. With this definition it is easy to show that the set of states $S ::
\Tset{\alpha~\var{c-state}}$ is finite.
\[ 
\var{S} = \{ \var{Start} \} \cup
  \{ \var{Init}~j \mid j \in \jondos \setminus \colls \} \cup 
  \{ \var{Mix}~j \mid j \in \jondos \} \cup \{ \var{End} \}
\] 

Often we are interested in the jondo referenced by the current state. We
introduce $\var{jondo-of} :: \alpha~\var{c-state} \to \alpha$ returning the
jondo if we are in an initial or mixing state:
\[ \var{jondo-of}~s = \Kcase~s~\Kof~\var{Init}~j \Rightarrow j \mid \var{Mix}~j \Rightarrow j \]

\begin{figure} 
\begin{center}
\begin{tikzpicture}

  \draw (8, 1.5) circle (1ex) node [above left] {$J_1$} ;
  \draw (8, 1) circle (1ex) node [above left] {$J_2$} ;
  \draw (8, 0.5) circle (1ex) node [above left] {$J_3$} ;
  \draw (8, 0) circle (1ex) node [above left] {$J_4$} ;
  \draw (8, -0.5) circle (1ex) node [above left] {$J_5$} ;
  \draw (8, -1.0) circle (1ex) node [above left] {$J_6$} ;
  \draw (8, -1.5) circle (1ex) node [above left] {$J_7$} ;

  \foreach \i in { 9.5, 11, 12.5, 14, 15.5 }
  { \draw (\i, 2) circle (1ex) node [above left] {$J_1$} ;
    \draw (\i, 1.5) circle (1ex) node [above left] {$J_2$} ;
    \draw (\i, 1) circle (1ex) node [above left] {$J_3$} ;
    \draw (\i, 0.5) circle (1ex) node [above left] {$J_4$} ;
    \draw (\i, 0) circle (1ex) node [above left] {$J_5$} ;
    \draw (\i, -0.5) circle (1ex) node [above left] {$J_6$} ;
    \draw (\i, -1.) circle (1ex) node [above left] {$J_7$} ;
    \draw[fill=gray] (\i, -1.5) circle (1ex) node [above left] {$C_1$} ;
    \draw[fill=gray] (\i, -2) circle (1ex) node [above left] {$C_2$} ; } ;
  
  \draw (6.5, 0) circle (1ex) ;

  \draw (17, 0) circle (1ex) node [above left] {$S$};

  \foreach \i in { 7.1, 8.6, 10.1, 11.6, 13.1, 14.6, 16.1 }
    \draw[dashed] (\i, 3.5) -- (\i, -2.5) ;

  \node at (6.5, 3) {\var{Start}} ;
  \node at (7.7, 3) {\var{Init}} ;
  \node at (9.2, 3) {$\var{Mix}_0$} ;
  \node at (10.7, 3) {$\var{Mix}_1$} ;
  \node at (12.2, 3) {$\var{Mix}_2$} ;
  \node at (13.7, 3) {$\var{Mix}_3$} ;
  \node at (15.2, 3) {$\var{Mix}_4$} ;
  \node at (16.7, 3) {\var{End}} ;
      
  \draw[->, thick, rounded corners=1pt]
    (6.5, 0) -- (8, 1.5) -- (9.5, 0.5) -- (11, 1.5) -- (12.5, -1.) -- (14, 0.5) -- (15.5, -1.5) -- (17, 0);
  
\end{tikzpicture}
\end{center}
\caption{The established route $J_1 - J_4 - J_3 - J_5 - J_4 - C_1 - S$}
\label{fig:crowds-mc}
\end{figure}
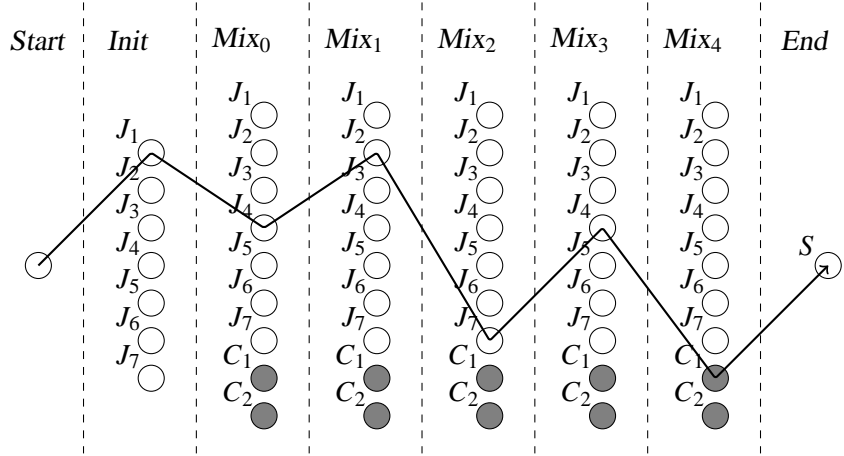

The transition matrix $\tau :: \alpha~\var{c-state} \to \alpha~\var{c-state} \to
\Real$ is defined by a case distinction on all possible transitions. The
probability for steps from \var{Start} are given by the distribution of the
initiating jondos \var{init}. The first routing jondo is arbitrarily chosen, and
the probability of going from a mixing state to a mixing state is the product of
$p_f$ to stay in the mixing phase and the probability $1 / J$ for the next
jondo. With probability $1 - p_f$ the mixing state is finished and than the
Markov chain stays in \var{End}. Figure~\ref{fig:crowds-mc} shows an example
path through the different phases.
\[ 
\begin{array}{@{}l@{~}r@{~}ll@{~}l}
\multicolumn{2}{@{}l}{\var{J} = | \jondos |} \\[1ex]
\multicolumn{2}{@{}l}{\var{H} = | \jondos \setminus \colls |} \\[1ex]
\tau~s~t = \Kcase~(s, t) & \Kof
 & (\var{Start}, & \var{Init}~j) & \Rightarrow \var{init}~j \\
 & \mid & (\var{Init}~j, & \var{Mix}~j') & \Rightarrow 1 / J \\
 & \mid & (\var{Mix}~j, & \var{Mix}~j') & \Rightarrow p_f / J \\
 & \mid & (\var{Mix}~j, & \var{End}) & \Rightarrow 1 - p_f \\
 & \mid & (\var{End}, & \var{End}) & \Rightarrow 1 \\
 & \mid & \_ & & \Rightarrow 0
\end{array}
\] 
This completes the definition of the Markov chain describing the route
establishment in the Crowds protocol. Finally we show that $S$ and $\tau$
describe a discrete-time Markov chain: 
\[ \Ktheorem \qquad \Kmarkovchain~S~\tau \]

\subsection{The jondo contacting the server is independent from the initiating jondo}

We define a number of path properties of our Markov chain. The functions
$\var{len} :: (\Nat \to \alpha~\var{c-state}) \to \Nat$, $\var{first-jondo} ::
(\Nat \to \alpha~\var{c-state}) \to \alpha$ and $\var{last-jondo} :: (\Nat \to
\alpha~\var{c-state}) \to \alpha$ operate on paths not containing the
$\var{Start}$ element. \var{len} returns the length of the mixing phase, i.e.\
how many $\var{Mix}$ states are in the path until $\var{End}$ is reached,
$\var{first-jondo}$ is the initiating jondo, and $\var{last-jondo}$ is the
jondo contacting the server.
\[ 
\begin{array}{@{}l}
\var{len}~\omega= (\Kleast~n.\;\omega~n = \var{End}) - 2 \\[1ex]
\var{first-jondo}~\omega = \var{jondo-of}~(\omega~0) \\[1ex]
\var{last-jondo}~\omega = \var{jondo-of}~(\omega~(\var{len}~\omega + 1))
\end{array}
\] 

The path functions \var{len}, \var{first-jondo} and \var{last-jondo} are well
defined on almost every path. The paths in our Markov chain do not contain the
\var{Start} element, so the paths start with an \var{Init} state. Hence for
almost every path we know that the first element is an initiating state, then for
the next \var{len} elements we have mixing states, and finally a tail of
\var{End} states:  
\[ 
\textstyle
\begin{array}{@{}l}
\Klemma\qquad\var{AE}_\var{Start}~\omega.\;\omega \in \Nat \to S\\[1ex]
\Klemma\qquad\var{AE}_\var{Start}~\omega.\;\exists j \in \jondos \setminus \colls.\;\omega~0 = \var{Init}~j\\[1ex]
\Klemma\qquad\var{AE}_\var{Start}~\omega.\;\forall i \le \var{len}~\omega.\;\exists j \in \jondos.\; \omega~(i + 1) = \var{Mix}~j \\[1ex]
\Klemma\qquad\var{AE}_\var{Start}~\omega.\;\forall i > \var{len}~\omega.\; \omega~(i + 1) = \var{End}
\end{array} \] 
With this we can easily show that the jondo contacting the server is independent
from the initiating jondo:
\[
\begin{array}{@{}l@{}l}
\multicolumn{2}{@{}l}{\Klemma} \\
\multicolumn{2}{@{}l}{\qquad\Kassumes~ l \in \jondos ~\Kand~ i \in \jondos \setminus \colls} \\
\qquad\keyword{shows}~ & \Pr(\omega.\; \var{first-jondo}~\omega = i \land
 \var{last-ncoll}~\omega = l) = \\
  & \Pr(\omega.\; \var{first-jondo}~\omega = i) *
  \Pr(\omega.\; \var{last-ncoll}~\omega = l)
\end{array}
\]

\subsection{Probability that initiating jondo contacts a collaborator}

The attacker model assumes that the collaborators want to detect the
initiator of a route. This is obviously only possible if one of the
collaborators is chosen as one of the mixing jondos.
We have two goals: (1)~If the numbers of collaborators is small, the
probability to contact a collaborator should be near zero. (2)~We want to
analyse the probability that the initiating jondo directly
contacts a collaborator. When we know the ratio of collaborators to jondos,
how can we adjust $p_f$, so that this probability is less or equal to $1 / 2$?

The random variable $\var{hit-colls} :: (\Nat \to \alpha~\var{c-state}) \to
\Bool$ is true if a collaborator participates in the mixing phase,
$\var{first-coll} :: (\Nat \to \alpha~\var{c-state}) \to \Nat$ is the mixing
phase in which the collaborator is hit, and $\var{last-ncoll} :: (\Nat \to
\alpha~\var{c-state}) \to \alpha$ is the last non-collaborating jondo, i.e.\ the
jondo contacting a collaborator.
\[ 
\begin{array}{@{}l}
\var{hit-colls}~\omega = \exists n, j \in \colls.\; \omega~n = \var{Mix}~j \\[1ex]
\var{first-coll}~\omega = (\Kleast~n.\;\exists j \in \colls.\;\omega~n = \var{Mix}~j) - 1 \\[1ex]
\var{last-ncoll}~\omega = \var{jondo-of}~(\omega~(\var{first-coll}~\omega))
\end{array}
\] 

The property we want to check only makes sense if a collaborator
participates in the mixing phase. So we first prove the probability to hit a
collaborator:
\[ \textstyle \Klemma~\Pr_\var{Start}(\omega.\; \var{hit-colls}~\omega) = (1 - H / J) / (1 - H / J * p_f) \]
We already see that the probability to hit a collaborator goes to $0$ if the
number of collaborators and $p_f$ stay constant and $J \longrightarrow \infty$.
Then $H / J \longrightarrow 1$ and hence
$\Pr_\var{Start}(\omega.\; \var{hit-colls}~\omega) \longrightarrow 0$. Thus our
first goal is satisfied.

Additionally, we want to control the probability that the initiating jondo hits
a collaborator. For this, we compute the probability to have a fixed first and
last non-collaborating jondo before we hit a collaborator:
\[
\begin{array}{@{}l@{}l}
\multicolumn{2}{@{}l}{\Klemma~\quad\text{P-\var{first-jondo}-\var{last-ncoll}}:} \\
\multicolumn{2}{@{}l}{\qquad\Kassumes~ l \in \jondos \setminus \colls ~\Kand~ i \in \jondos \setminus
  \colls} \\
\qquad\keyword{shows}~ & \Pr(\omega.\; \var{first-jondo}~\omega = i \land
 \var{last-ncoll}~\omega = l \mid \var{hit-colls}~\omega) = \\
  & \init~i * (p_f / J + (\Kif~ i = l ~\Kthen~ 1 - H / J * p_f ~\Kelse~ 0))
\end{array}
\]
Note that the conditional probability does not divide by $0$ because $\Pr_\var{Start}(\omega.\; \var{hit-colls}~\omega) \neq 0$ by the previous lemma.
By summing up over all possible non-collaborating jondos we show the probability
that the last non-collaborating jondo is the initiating jondo:
\[ \textstyle \Ktheorem~\Pr_\var{Start}(\omega.\; \var{first-jondo}~\omega = \var{last-ncoll}~\omega \mid \var{hit-colls}~\omega) = 1 - (H - 1) / J * p_f \]
With this we can now enforce that the probability that the initiating jondo hits
a collaborator is less or equal to $\frac{1}{2}$:
\[
\begin{array}{@{}l@{}l}
 \textstyle \Klemma\qquad H > 1 ~\land~ J / (2 * (H - 1)) \le p_f \implies \\
\quad
\Pr_\var{Start}(\omega.\; \var{first-jondo}~\omega = \var{last-ncoll}~\omega \mid \var{hit-colls}~\omega) \le \frac{1}{2}
\end{array}
\]
Reiter and Rubin~\cite{reiter1998crowds} call this probably innocent.
Because $p_f < 1$ this is only possible if $1 / 2 < (H - 1) / J$,
i.e.\ more than half of the jondos are non-collaborating. This meets our
second goal.

\subsection{Information gained by the collaborators}


Obviously, in Isabelle/HOL we are not only restricted to state probabilities or
expectations. For example, for quantitative information flow analysis, similar
to the analysis by Malacaria~\cite{malacaria2007securitythreats}, we are
interested in the mutual information $\mathcal{I}_s(X ; Y)$ between two random
variables $X$ and $Y$. The mutual information is formalized in Isabelle/HOL
using the \RN\ derivative. However, we know that if $X$ and $Y$ are simple
functions, i.e.\ functions with a finite range, then $\mathcal{I}_s(X ; Y)$ can
be computed in the known discrete way:
\[ 
  \textstyle
  \begin{array}{@{}l}
  \Klemma\qquad\var{simple-function}_s~X \implies \var{simple-function}_s~Y \implies \\
  \quad \mathcal{I}_s(X ; Y) = \sum_{(x,y) \in \{ (X x, Y x) \mid x. x \in \Omega \}}. 
    \Pr_s(\omega.\; X~\omega = x \land Y~\omega = y) *
    \\
   \qquad
    \log_2 \big(\Pr_s(\omega.\; X~\omega = x \land Y~\omega = y) /
     (\Pr_s(\omega.\; X~\omega = x) * \Pr_s(\omega.\; Y~\omega = y))\big)
   \end{array}
\] 
We are only interested in runs which hit a collaborator. To use mutual
information with this restriction we introduce the conditional probability
$\Pr_\var{hit-colls}$, with the condition that each run hits a collaborator.
Its characteristic property (we omit the technical definition) is
\[ 
  \textstyle
  \Klemma\qquad \var{measurable}_s~P \implies \Pr_\var{hit-colls}(\omega.\; P~\omega) =
    \Pr_\var{Start}(\omega.\; P~\omega \mid \var{hit-colls}~\omega)
\] 
With this property and lemma P-\var{first-jondo}-\var{last-ncoll} we
can now show an upper bound for the information flow: 
\[ 
\textstyle
\begin{array}{@{}l}
  \Ktheorem\qquad\mathcal{I}_\var{hit-colls}(\var{first-jondo}; \var{last-ncoll})
    \le (1 - (H - 1) / J * p_f) * \log_2 H
\end{array}
\] 
This supports the intuitive understanding that the information the attackers can
gain is restricted by the probability that the initiating jondo is the jondo
directly contacting a collaborator.

\section{Related Work}

There is already some work to verify parametric probabilistic models.
Hermanns~\emph{et~al.}~\cite{hermanns2008pcegar} implement a probabilistic
variant of counterexample-guided abstraction refinement (CEGAR). They handle
infinite state spaces by breaking them up into finite partitions.
Hahn~\emph{et~al.}~\cite{hahn2011parametricmc} allows parametric transition
probabilities. The number of states is still fixed, but the transition
probabilities are rational functions over parameter variables.
Katoen~\emph{et~al.}~\cite{katoen2010invariant} present a method to generate and
use quantitative invariants for linear probabilistic programs. Their motivation
is to use these invariants to augment interactive proofs.


Now we survey other work that models probabilistic systems in an interactive
theorem prover.

We build directly on the formalization of Markov chain theory developed for
our verification of pCTL model checking~\cite{hoelzl2012verifyingpctl}, which
builds on a formalization of measure theory~\cite{hoelzl2011measure}.  Ultimately,
all of the work cited in this section builds on the work of Hurd (see below).
However, instead of Hurd's probability space $\Nat\to\Bool$ we have a
probability space on arbitrary functions. This allows for a natural
formalization of Markov chains over arbitrary state spaces and needs no
encoding into booleans.

The formalization of probability theory in HOL starts with Hurd's
thesis~\cite{hurd02thesis}. He introduces measure theory, proves Caratheodory's
theorem about the existence of measure spaces and uses it to introduce a
probability space on infinite boolean sequences. He defines concrete random
variables with Bernoulli or uniform distribution. Using this work he also
analyses a symmetric simple random walk.
Hasan~\emph{et~al.}~\cite{hasan2009expectation} formalize the analysis of
continuous random variables on Hurd's probability space. However, their work is
quite different from ours in that they do not employ Markov chains.
Based on Hurd's work, Liu~\emph{et~al.}~\cite{liu2011markovchains} define when
a stochastic process is a Markov chain. Their theory does not provide
everything we need: it is restricted to stochastic processes on Hurd's
probability space $\Nat\to\Bool$ and does not construct the path space of Markov
chains defined by transition probabilities.
Coble~\cite{coble09thesis} formalizes information theory on finite probability
spaces. He applies it to a quantitative information flow analysis of the
Dining~Cryptographers protocol. 
Hurd~\emph{et~al.}~\cite{hurd05pgcl} in HOL4 and Audebaud and
Paulin-Mohring~\cite{audebaud2009randomizedalgos} in Coq formalize semantics
of probabilistic programs. Both reason about the probability of program
termination and only allow discrete distributions for the result values.

\section{Conclusion}

The formalizations are available in the
Archive~of~Formal~Proofs~\cite{hoelzl2012afp-markov}. For the ZeroConf protocol
the formalization was done in a couple of days and required
approx.\ 260~lines of Isabelle/HOL theory. The Crowds protocol requires approx.\
1060~lines of Isabelle/HOL theory and it took one person a couple of weeks to
verify. The time necessary for the verification includes finding an estimation
for the information gained when a collaborator is hit.
The probabilities we verified for the ZeroConf protocol and the Crowds protocol
are expressible as PCTL formulas. However this is not a restriction of
Isabelle/HOL. We can express $\omega$-regular properties or multiple reward
structures easily in higher-order logic.

Our future goals include more powerful models like Markov decision processes and
continuous-time models but also the certification of probabilistic model checker
runs in Isabelle/HOL.

\section*{Acknowledgment}

We thank Sergio Giro for reading and commenting on a draft of this paper. We
also thank the anonymous reviewers for the references on parametric
probabilistic model checking.

\bibliographystyle{eptcs}
\bibliography{QFM2012.bib}
\end{document}